\documentstyle[11pt]{article}

\begin{document}


\begin{center}
{\Large\bf Building a case for a
Planck-scale-deformed boost action:\\
the Planck-scale particle-localization limit\footnote{This essay
received an honorable mention in the 2005 Essay Competition of
the Gravity Research Foundation}}
\end{center}

\vskip 0.5 cm
\begin{center}
{\bf Giovanni AMELINO-CAMELIA}\\
\end{center}

\begin{center}
{\small {\it Dip.~Fisica Univ.~Roma ``La Sapienza''
and Sez.~Roma1 INFN, }\\
{\it Piazzale Moro 2, Roma, Italy}\\
amelino\@roma1.infn.it}
\end{center}

\vskip 0.5 cm
\begin{center}
{\bf ABSTRACT}
\end{center}

\begin{quotation}
\leftskip=0.6in \rightskip=0.6in

{\small
\noindent
 ``Doubly-special relativity" (DSR), the idea of a Planck-scale Minkowski limit
 that is still a relativistic
 theory, but with both the Planck scale and
the speed-of-light scale as nontrivial relativistic invariants,
was proposed (gr-qc/0012051) as a physics intuition for several scenarios
which may arise in the study of the quantum-gravity problem,
but most DSR studies focused exclusively
on the search of formalisms for the description of a specific example
of such a Minkowski limit.
A novel contribution to the DSR physics intuition came from a
recent paper by Smolin (hep-th/0501091)
suggesting that the emergence of the Planck scale as a second
nontrivial relativistic invariant might be inevitable in
quantum gravity, relying only on some rather robust expectations concerning
the semiclassical approximation of quantum gravity.
I here attempt to strengthen Smolin's argument by observing
that an analysis of some independently-proposed
Planck-scale particle-localization limits, such as the ``Generalized
Uncertainty Principle" often attributed
to string theory in the literature, also suggests that the
emergence of a DSR Minkowski limit might be inevitable.
I discuss a possible link between this observation and recent
results on logarithmic corrections to the entropy-area black-hole formula,
and I observe that both the analysis here reported and Smolin's analysis
appear to suggest that the examples of DSR Minkowski limits for which
a formalism has been sought in the literature might not be sufficiently
general.
I also stress that, as we now contemplate
the hypothesis of a DSR Minkowski limit,
there is an additional challenge for those in the quantum-gravity community
attributing to the Planck length the role of ``fundamental length scale".
}
\end{quotation}

\baselineskip 14pt plus .5pt minus .5pt

\subsubsection*{Beyond Special Relativity}
One might say that special relativity went into retirement at a very
young age, since the advent of general relativity, in 1916,
took away from special relativity the status of ``fundamental" ingredient
of the laws of Nature. But actually special relativity, now 100 years old,
continues
to work hard for physicists. In quite a few contexts involving gravitational
phenomena one gets away describing the spacetime metric $g_{\mu \nu}$
in terms of a Minkowski background metric, $\eta_{\mu \nu}$,
and a ``gravity field" Lorentz tensor $h_{\mu \nu}$,
related to $g$ and $\eta$ by the
relation $h_{\mu \nu} = g_{\mu \nu} - \eta_{\mu \nu}$.
And special relativity still reigns supreme in the vast
class of phenomena studied in particle physics, where
one can safely assume that the processes
unfold in a Minkowski background spacetime, with
metric $\eta_{\mu \nu}$, and that ``gravitational
interactions" among particles are negligible.
Special relativity still holds an essentially fundamental status within
particle physics.
This longevity is due mostly to the fact that
special relativity, initially introduced to address some issues of concern
in classical mechanics (notably the form of Maxwell equations),
turned out to deal rather well with the new structures
of quantum mechanics. Special relativity
and quantum mechanics coexist rather peacefully in the framework provided
by relativistic quantum field theory.
But several authors have argued that the transition from
(special-) relativistic quantum field theory
to (the still unknown) ``quantum gravity"
might force special relativity  to relinquish
even its present privileged status within particle physics, as soon as
we acquire sensitivity to Planck-scale
corrections to particle-physics processes.

The description of Planck-scale
corrections to particle-physics processes will be
a key aspect of the Minkowski limit of quantum gravity.
In our current conceptual framework
special relativity emerges in the Minkowski limit,
where one deals with situations that allow
the adoption of a Minkowski metric throughout,
and one might wonder whether the Minkowski limit of quantum gravity
could still be governed by special relativity.
The issue will be of particular interest if quantum gravity
admits a limit
in which one can
assume throughout a (expectation value of the) metric of
Minkowski type, but some Planck-scale features
of the fundamental description of spacetime
(such as spacetime discreteness and/or spacetime noncommutativity)
are still not completely negligible.
I will denominate ``nontrivial Minkowski limit" this type of
Minkowski limit in which essentially the role of the Planck scale
in the description of gravitational interactions (expressing the
gravitational constant $G$ in terms of the Planck scale)
can be ignored, but the possible role of the Planck scale in
spacetime structure/kinematics is still significant.
It is of course not obvious that the correct quantum gravity should
admit such a nontrivial Minkowski limit. With the little we presently
know about the quantum-gravity problem we must be open to the possibility
that the Minkowski limit would actually be trivial, {\it i.e.}
that whenever the role of the Planck scale
in the description of gravitational interactions can be neglected
one should also
neglect the role of the Planck scale in spacetime structure/kinematics.
But the hypothesis of a nontrivial Minkowski limit is worth exploring:
it would be ``extremely kind" of quantum gravity to admit such a limit,
since it might open a wide range of opportunities for
accessible experimental
verification (see, {\it e.g.},
Refs.~\cite{grbgac,gampul,kifu,ita,gactp,jaco,nickcpt}).

For various approaches to the quantum-gravity problem evidence
as emerged in support of this
possibility of a nontrivial Minkowski limit.
While there is no fully-developed proposed solution of the quantum-gravity
problem based on a fundamentally noncommutative spacetime picture,
it has been observed that the
hypothesis that in  general the correct fundamental description of
spacetime should involve noncommutativity can imply  that in particular
the Minkowski limit is described in terms of noncommuting spacetime coordinates,
and this is
found~\cite{majrue,kpoinap,Monaco,susskind,dougnekr,alessfranc,bertoNC}
to naturally lead to a nontrivial Minkowski limit with departures
from classical Poincar\'e symmetry.
In the literature on the loop-quantum-gravity approach
one finds a large number (although all of preliminary nature)
of arguments~\cite{gampul,mexweave,thiemLS,leeDispRel,kodadsr}
supporting the possibility of
a nontrivial Minkowski limit, primarily characterized by a
Planck-scale-modified energy-momentum (dispersion) relation.
For the string-theory approach, while there are no studies arguing that
the availability of a nontrivial Minkowski limit is necessary,
there is a large literature (see, {\it e.g.},
Refs.~\cite{susskind,dougnekr}
and references therein) on a nontrivial Minkowski limit with broken
Lorentz symmetry.

The fate of Poincar\'e symmetry in such nontrivial Minkowski
limits is of course a key issue both phenomenologically and from
a conceptual perspective.
In the large number of studies produced between 1997 and 2000 on the
possibility of a nontrivial Minkowski limit it was always assumed
that Poincar\'e (and in particular Lorentz) symmetry would be broken:
the Galilei Relativity Principle would not hold with Planck-scale
accuracy.
On the basis of an analogy with the century-old process which
led from Galilei/Newton Relativity, through the analysis of Maxwell's
electrodynamics (first viewed as a manifestation of
a ether violating the relativity principle, but ultimately understood
as a manifestation of a needed transition in the formulation
of the relativistic theory), to Einstein's Special Relativity,
I argued in Ref.~\cite{gacdsr} that the Minkowski limit of quantum
gravity might be characterized by a ``doubly special relativity" (DSR),
a relativistic theory with two, rather than one,
nontrivial relativistic invariants (the Planck scale in addition
to the speed-of-light scale), but still fully compatible with
the Galilei Relativity Principle.

At present this DSR proposal is still confined in the limbo
of a physics scenario for which good mathematics (or at least a fully
satisfactory use of mathematics) is being sought.
The prototypical example of a quantum-gravity theory that would
need the DSR idea is a quantum-gravity theory in which in the Minkowski
limit one finds that the Planck (length) scale sets an observer-independent
minimum allowed value of wavelength.
But of course there is a pletora of other requirements
which would could be advocated in order to find a DSR Minkowski limit.
The abundance of possible physical principles that one might consider
from a DSR perspective is counter-acted by an apparently scarce
ensamble of mathematical formalisms that could be used
to describe such a DSR Minkowski limit.
For example attempts based on the $\kappa$-Minkowski noncommutative spacetime,
and an associated Hopf-algebra
structure~\cite{majrue,kpoinap,alessfranc,jurekdsr},
provide a description which, while exposing some structures that
are of potential relevance for the DSR idea,
at present is still unsatisfactory
as far as interpretation of the formalism in the one-particle
sector, and both mathematically unsettled and interpretationally
unclear for multiparticle systems~\cite{mg10qg5}.
It is legitimate to hope that an improved scheme based on this
Hopf-algebra mathematics might turn out to circumvent these limitations,
but some key difficulties must be overcome.
Some arguments suggesting a possible relevance of the DSR proposal
for the Minkowski limit of loop quantum gravity have
been proposed~\cite{leedsr,kodadsr}, but in order to render
these arguments more precise it would of course be first necessary
to establish what procedure actually leads to the
Minkowski limit in loop quantum gravity.

\subsubsection*{A gravity rainbow?}
One of the many attempts to introduce mathematical structures
suitable for the description of a DSR Minkowski limit
is based on the introduction of an energy-dependent metric.
This technique was named ``gravity rainbow" by Magueijo and Smolin,
who first proposed it as a tool for DSR research~\cite{leedsr}.
It is probably fair to say that also this proposal still lacks
a fully developed interpretation and implementation.
It is easy to see however how the language of an energy-dependent
metric could be used for an effective description of modified
energy-momentum relations in the Minkowski limit.
For example, if the modified dispersion relation is of the type
\begin{equation}
{\cal C} = p^2 + f(E,E_p) E^2
~,
\label{rainb1}
\end{equation}
where $C$ is the deformed ``mass Casimir" and $E_p$ denotes
the Planck energy scale,
one could introduce an energy-dependent
metric $g_{\mu \nu}(E,E_p)$ such that  $g_{0 0}(E,E_p)= f(E,E_p) \eta_{0 0}$
and $g_{i j} = \eta_{i j}$
and describe the same dispersion relation as
\begin{equation}
{\cal C} = p^\mu g_{\mu \nu}(E,E_p) p^\nu
~.
\label{rainb2}
\end{equation}

The possibility of an observer-independent Planck-scale modification
of the energy-momentum dispersion relation is one of the most studied
possibilities as a ``physics ingredient" for a DSR Minkowski limit,
and therefore the gravity rainbow formulation might indeed be relevant
for DSR research.
However, it might be harder to formalize simply
as an energy-dependent metric
other aspects of a DSR Minkowski limit (other than the dispersion relation).
The dispersion relation is by construction a link between the energy
of a given particle and its momentum, so it is clear
in that context to what energy one should should calibrate the metric,
but in other contexts, especially when several energy scales are
involved, a simple-minded implementation of an energy dependence
of the metric might lead to ambiguities.
I shall get back to this issue at a later point of the essay.

\subsubsection*{Motivation for DSR from a Planck-scale particle-localization limit}
The key point I intend to make in this essay concerns a perspective
on some much-discussed Planck-scale particle-localization limits
that also suggests, in the same sense of the argument presented by Smolin
in Ref.~\cite{leeDSRalways}, that the availability of a DSR Minkowski
limit might indeed be inevitable.

I intend to focus on the relation that describes
the minimum uncertainty $\delta x$
in the position of a particle of given (expectation value of) energy $E$.
In relativistic quantum field theory
it is well established that~\cite{landau,bj1e2}
\begin{equation}
\delta x \ge \frac{1}{E}
~,
\label{undeform}
\end{equation}
and this relation is a key aspect
of the interplay
between (and ultimately responsible for the compatibility of)
special relativity and quantum mechanics.
If one wanted to localize
a particle ${\cal P}$ of mass $m$, in a frame where it is at rest,
with accuracy better than $1/m$,
the procedure should necessarily
(because of the Heisenberg uncertainty principle)
involve an exchange of energy greater than $m$
between the probe used in the procedure and ${\cal P}$.
But the availability of energy greater than $m$
would be sufficient (according to special relativity)
to create additional copies of ${\cal P}$, thereby rendering
the procedure results not meaningful~\cite{landau,bj1e2}
as a localization of ${\cal P}$.
In the rest frame of the particle
one therefore finds the following localization limit
\begin{equation}
\delta x_{rest} \ge  \frac{1}{m}
~,
\label{undeformrest}
\end{equation}
where $\delta x_{rest}$ denotes the uncertainty in any of the three spatial
coordinates $\delta x_{1,rest}$, $\delta x_{2,rest}$, $\delta x_{3,rest}$.
Since the $\delta x_{i}$ are lengths (the size of the ``uncertainty interval"
on the $x_i$ axis), and are therefore subject to FitzGerald-Lorentz (boost)
contraction, if indeed in its rest frame the particle is confined to a volume of
size $1/m$, then an observer moving (with respect to the particle's rest frame)
with speed $V$ along the ``$x_1$" direction would attribute to
the particle
a $\delta x_1 = \delta x_{1,rest} \sqrt{1-V^2} = \sqrt{1-V^2}/m = 1/E$,
while $\delta x_2 =\delta x_3 =1/m$.
So clearly (\ref{undeform}) must hold in any frame.

Whereas in nonrelativistic quantum mechanics one characterizes completely
the limitations on particle localization  through
the Heisenberg relation
\begin{equation}
\delta x \ge  \frac{1}{\delta p}
~,
\label{heiseold}
\end{equation}
and there is still no absolute limitation on
the accuracy of localization of a particle
of any energy $E$ (if one accepts a correspondingly large
uncertainty in the momentum of the particle),
in relativistic quantum mechanics (quantum field theory) one
must take into account (as stressed above)
the possibility of particle production, which
introduces a further (and absolute) limitation,
codified by (\ref{undeform}), on the accuracy
reachable in localization of a particle of energy $E$.

The introduction of special-relativistic effects within the conceptual
framework of quantum mechanics leads to an absolute limit on the
localization of a particle of energy $E$, but still allows the
abstraction of a particle that can be sharply localized, in
the $E \rightarrow \infty$ limit.
If one then also introduces general-relativistic effects,
according to an intuition which finds support in a large literature
 (see, {\it e.g.}, Refs.~\cite{meadpadma,venegross,ahlu1994,ng1994gacmpla,garay}),
the abstraction of sharp localization of a particle should be completely
removed. Essentially the classical concept of a spacetime point would
loose operative meaning, since no particle can ever
provide ``physical identity" to that spacetime point.

The expectation that (\ref{undeform}) should fail at the Planck scale
is based on the fact that the Planck length appears to set
an absolute limit on the localization
of a particle. According to (\ref{undeform}) one could measure the position
of a ultrahigh-energy particle with corresponding ultrahigh precision, and
for particles of sufficiently high energy one could achieve even localization
better than the Planck length.
Several arguments~\cite{meadpadma,venegross,ahlu1994,ng1994gacmpla,garay}
suggest that sub-Planckian localization accuracy
should be impossible. The simplest such argument
observes that in order to localize a particle with accuracy $\delta x$,
such that $\delta x < L_p$ (where $L_p$ denotes the Planck length),
there should be a stage in the measurement procedure in which the probe
and the particle whose position is being measured exchange energy
greater than the Planck scale localized in
a region of size smaller than the Planck length.
This should lead to the formation of
a black hole, which then prevents the localization procedure from completing
successfully.

The intuition that sub-Planckian localization cannot be possible
is shared by a large majority of the quantum-gravity community,
but there has been no previous discussion of possible
implications for equation (\ref{undeform}),
and the issue of consistency between the familiar form of Lorentz boosts
and the Planck-length localization limit was never addressed.
Most attempts of introducing formal structures that would reflect
the Planckian localization limit were confined (sometimes implicitly)
to the context
of Planck-scale modifications of nonrelativistic
quantum mechanics. In
particular several papers (see, {\it e.g.}, Ref.~\cite{kempmang})
have used the Planck-length limit
as motivation for the study of some Hilbert-space pictures
that would be consistent with the existence of a minimum length.

I propose that any fruitful characterization of Planck-scale localization
limits should include an analysis of the corresponding modifications of
the relation (\ref{undeform}). This is rather clear if one thinks of relativistic
quantum field theory as the starting point for the search of quantum gravity,
and it also reflects the fact that (\ref{undeform}) is the ultimate
uncertainty principle for localization in our current classical-spacetime
theories, and therefore,
as we seek, with quantum gravity, a new description of spacetime structure,
modifications of (\ref{undeform}) will provide the most intuitive
characterization of possible nonclassical features in spacetime geometry.
I will consider modifications of (\ref{undeform})
such that the Planck-length localization limit is enforced,
and it will become quickly clear that such modifications inevitably require
departures from at least some special-relativistic laws.
Let us consider first the possibility that boost still act
in the familiar Lorentz way on energy, so that the relation between
the energy $E$ in our chosen frame and the rest energy $m$
is still given by $E=m/\sqrt{1-V^2}$, and energy is still
unbounded from above.
In this case the only way to ensure $\delta x \ge L_p$
is to replace (\ref{undeform}) with a formula of the
type $\delta x \ge f(E;L_p)$ where the function $f$ has minimum
value $L_p$.
But then of course boost-contraction of $\delta x$
(the relation between $\delta x$ and $\delta x_{rest}$)
should be accordingly modified.
For an explicit illustrative
example of this possibility let us consider a simple ``see-saw formula"
\begin{equation}
\delta x \ge \frac{1}{E} + \frac{L_p^2}{4} E
~,
\label{new1}
\end{equation}
which is inspired by the ``Generalized Uncertainty Principle"
($\delta x \ge \frac{1}{\delta p} + \alpha \delta p$)
that is often attributed to string theory in the
literature~\cite{venegross,kempmang}.
The added term $L_p^2 E/4$ ensures that $\delta x$
can never be smaller than $L_p$.
If $E$ still transforms under boosts
in the special-relativistic way, then the covariance of (\ref{new1})
would require a Planck-scale-deformed boost contraction of lengths
such that
\begin{equation}
\delta x = \frac{\delta x_{rest}+\sqrt{\delta x_{rest}^2-L_p^2}}{2} \sqrt{1-V^2}
+  \frac{ L_p^2}{2 \left(\delta x_{rest}+\sqrt{\delta x_{rest}^2
- L_p^2} \right) \sqrt{1-V^2}}
\label{new2}
\end{equation}
(with an accompanying requirement that $\delta x_{rest} \ge L_p$).

If instead we allow a Planck-scale-deformed action of boosts on energy,
then actually equation (\ref{undeform}) could even preserve its original form,
as a relation between $\delta x$ and $E$ (but necessarily changing form
as a relation between $\delta x$ and the rest mass of the particle $m$).
To give once again an illustrative example, let me consider
the following Planck-scale-modified boost relation between the energy
in the chosen frame and the rest mass
\begin{equation}
E= \frac{1}{L_p} \tanh \left( \frac{L_p m}{\sqrt{1-V^2}} \right)
~,
\label{new3}
\end{equation}
which, as a result of familiar properties of the hyperbolic tangent,
introduces a maximum energy $1/L_p$ for fundamental particles
(but as long as $E \ll 1/L_p$ is in excellent agreement with $E=m/\sqrt{1-V^2}$).
If this deformed boost relation between $E$ and $m$
was assumed, equation (\ref{undeform}) would then
automatically be consistent with the requirement $\delta x \ge L_p$.

Of course, as long as the only hint we have is the $\delta x \ge L_p$ relation,
we cannot establish whether (\ref{undeform}) should be
modified explicitly, as in the illustrative example (\ref{new2}),
or only implicitly, because of a change in the relation
between $E$ and $m$, as in the illustrative example (\ref{new3}).
Alternative candidate solutions of the quantum-gravity problem
might actually lead to different modifications of relation (\ref{undeform}).
But we have seen that in any case
a Planck-scale deformation of the action of boosts on
at least some observables is required.
If indeed the Minkowski limit of quantum gravity is affected by
a Planck-scale localization limit,
and if this emerges without spoiling the Galilei Relativity Principle,
one should then conclude that this Minkowski limit is
a DSR Minkowski limit.

\subsubsection*{Encouragement from black-hole entropy results?}
Since the perspective I am adopting is the one of building a case
for a DSR Minkowski limit on the basis of the argument reported by
Smolin in Ref.~\cite{leeDSRalways} and on the basis of my description
of a Planck-scale particle-localization limit, it is useful to
stress that a Planck-scale particle-localization limit
can also be independently motivated (as implicitly argued in
Refs.~\cite{bh1e2,bh3}) on the basis of the
expectation~\cite{stringbek,lqgbek}
of log corrections to the area-entropy relation for black holes.
I can quickly reach this conclusion by revisiting
the Bekenstein
argument~\cite{bek} for the area-entropy relation,
using (\ref{new1}) in place of $\delta x \ge 1/E$.

As in the original Bekenstein
argument~\cite{bek}, I take as starting
point the general-relativity result~\cite{christo} which
establishes that the area of a
black hole changes according to $\Delta A \ge 8 \pi L_p^2 E s $
when a classical particle of energy $E$ and size
$s$ is absorbed. Whereas Bekenstein
describes the size of the
particle in terms of the uncertainty in its position
according to $s \simeq \delta x \simeq 1/E$,
I shall assume $s \simeq \delta x \simeq 1/E + L_p^2 E/4$,
thereby obtaining
\begin{eqnarray}
\Delta A \ge \alpha L_p^2  - \beta \frac{L_p^4}{(\delta x)^2}
~.
\label{minDagacm1}
\end{eqnarray}
where $\alpha$ and $\beta$ are numerical coefficients whose precise
value is irrelevant for my point.
I then use~\cite{bh1e2,bh3} the fact that in falling in the black hole
the particle acquires~\cite{calib2,smaller,bigger}
position uncertainty $\delta x \sim R_S$, where $R_S$ is the
Schwarzschild radius (and
of course $A = 4 \pi R_S^2$) to obtain
\begin{eqnarray}
\Delta A \ge
\alpha L_p^2 - \beta \frac{4 \pi L_p^4}{A}
~.
\label{minDagac}
\end{eqnarray}
From (\ref{minDagac}) one derives an area-entropy relation
assuming (again following Bekenstein)
that the entropy of the black hole depends only on its area and
that the minimum increase of entropy should be,
independently of the value of the area, $\ln 2$:
\begin{equation}
\frac{dS}{dA} \simeq \frac{min (\Delta S)}{min (\Delta
A)} \simeq  \frac{\ln 2}{\alpha L_p^2 - \beta \frac{4 \pi L_p^4}{A}}
~, \label{minDa}
\end{equation}
which gives (up to an irrelevant constant contribution to entropy)
\begin{equation}
S \sim \alpha' \frac{A}{L_p^2} + \beta' \ln \frac{A}{L_p^2}
~. \label{entropy1}
\end{equation}
Again for notational convenience I introduced numerical
coefficients $\alpha'$ and $\beta'$. These coefficients
are anyway not reliably computed using the Bekenstein
argument.
Of course, the original version of the Bekenstein
argument, using only $\delta x \ge 1/E$, only includes the
leading-order term, {\it i.e.} a linear relation between
entropy and area.
The famous result $\alpha' = 1/4$ for the coefficient of
this linear term was first obtained
by Hawking~\cite{hawkBH}.
Improving the Bekenstein
argument by using the Planck-scale
particle-localization limit $\delta x \geq 1/E + L_p^2 E/4$
one finds also a log correction to the entropy-area relation,
and this is indeed encouraging since
both in String Theory and in Loop Quantum Gravity
it has been argued that such log corrections should be present.
Looking at this from a perspective bottom-to-top one could say
that the arguments providing support for log corrections
to the entropy-area relation also provide indirect support
for a Planck-scale particle-localization limit of
the type (\ref{new1}).

\subsubsection*{A new avenue for DSR research}
The Doubly-special relativity idea was proposed~\cite{gacdsr} as an intuition
for an open physics problem (rather than for a certain mathematical formalism):
the Minkowski limit of quantum
gravity should be described by a relativistic theory (a theory compatible
with the Galilei Relativity Principle)
with both the  Planck scale and
the speed-of-light scale as nontrivial relativistic invariant
(scales which enter in the formulas that connect different observers).
However, most of the attempts trying to introduce formalisms
suitable for the description of a DSR Minkowski limit focused
on the possibility that the modification of the laws of transformation
between inertial observers would be such that space rotations
and boosts should be described in terms of nonlinear realizations
of the classical Lorentz symmetry group.
The relevant work was so focused on this possibility that
many readers from outside the ``DSR community" have identified
the concept of a ``DSR Minkowski limit" and the mathematics of
nonlinear realizations
of the classical Lorentz symmetry group.
Moreover, it is assumed that the classical Lorentz symmetry group
should be realized nonlinearly in exactly the same way
throughout the formalism (or at least on all aspects
of the energy-momentum-space).

However, if one looks at the type of structures that quantum gravity
might impose on its Minkowski limit one easily finds encouragement
to look into other possible mathematical formalisms.
This is in particular true of the Magueijo-Smolin ``gravity rainbow"
proposal for a DSR Minkowski limit and also for the argument I gave
above based on
a Planck-scale localization limit.

Let us first consider the possibility of a
gravity-rainbow-type DSR Minkowski limit.
As stressed above, the abstraction of an energy-dependent metric
provides a rather intuitive description of Planck-scale
modifications of the energy-momentum dispersion relation,
but attempts to apply the generic concept of an
energy-dependent metric to other physically meaningful quantities,
especially when several energy scales are
involved, might easily run into ambiguities\footnote{Actually, it is well known
that in some DSR schemes, even some of those considered in some of the earliest
DSR studies~\cite{gacdsr,dsrIJrev},
the abstraction of a spacetime (and of a metric)
is not available in general.
As stressed in Ref.~\cite{dsrIJrev},
if an observer-independent nonlinear deformation of the dispersion relation
is adopted and velocity is described by $v = dE/dp$, then one
observer, ``$O$", could see two particles with different masses $m_A$ and $m_B$
moving at the same speed and following the same trajectory (for $O$ particles $A$
and $B$ are ``near" at all times), while for a second observer $O'$
the same two particles would have different velocities (so they could not possibly
be ``near" at all times).
This then leads inevitably~\cite{dsrIJrev} to considering spacetime as an approximate
concept, only valid within a certain class of observations and with a certain
level of approximation. In the low-energy regime one could still introduce a
spacetime and an associated metric. And one could still have~\cite{dsrIJrev} a spacetime
and a metric (but a different metric, depending on the energy) in certain
special high-energy processes ({\it e.g.} processes involving all particles with
the same energy and mass). But in general ({\it e.g.} for processes involving several
particles with large hierarchies of energies) one should do without~\cite{dsrIJrev} a
conventional concept of spacetime
(and of course without a conventional concept of metric).}.
I intend to argue that a scheme which may not run into ambiguities even when
naively implemented is obtained
if one replaces the energy-dependent metric with a corresponding
statement of nonlinear relation between covariant fourmomentum
and contravariant fourmomentum.
After all (in an appropriate sense) the Minkowski limit does not
really require us to make explicit reference to a metric.
The ordinary $\eta_{\mu \nu}$ is only used to lower and raise indices,
and in particular it is used to relate (linearly) the
covariant fourmomentum
and the contravariant fourmomentum.
The energy dependence of the metric in the Minkowski limit
could be a simple way to express a requirement of nonlinear relation
between the covariant fourmomentum
and the contravariant fourmomentum.
One of the two (say, the covariant fourmomentum) could still transform
according to ordinary special relativity, but then the relativistic
properties of the other would codify departures from the special-relativistic
predictions.
This leads one to consider a previously unexplored possibility for the
construction of DSR Minkowski limits.
Whereas usually in DSR research one assumes that the same nonlinear
realization of the Lorentz symmetry group should be applied to
all energy-momentum-space quantities, one should perhaps also contemplate
the possibility that, say, the covariant fourmomentum
still transforms linearly under Lorentz transformations, while
the contravariant fourmomentum might indeed transform nonlinearly.

One can look from an analogous perspective at the point
concerning
a Planck-scale particle-localization limit which I made in this essay.
One of the scenarios I considered assumed that the energy $E$
still transforms linearly under Lorentz boosts, but the
position uncertainty $\delta x$ would transform nonlinearly
(and of course then the relation between $E$ and $\delta x$ should
be nonlinear).

In general one could notice that
in the classical Minkowski limit various quantities, such as
the covariant fourmomentum, the contravariant fourmomentum,
and the frequency/wavenumber fourvector, all transform in the same
linear way under Poincar\'e transformations,
but in a Planck-scale-accurate description of the Minkowski limit
some differences may arise, and, for example, the transformation rules
of some of these quantities might still be linear, while some other
of these quantities might transform nonlinearly.

\subsubsection*{On the criteria for a DSR Minkowski limit}
Part of the thesis presented in this essay is that the search of
formalisms suitable for the description of a DSR Minkowski limit
might have been too narrow. There is a variety of physical postulates
that one could consider for the Minkowski limit of a quantum-gravity theory,
and it appears likely that each of this possibilities might require
different mathematics for its description.
The tendency by some authors
to identify the ``physics project"
of a DSR Minkowski limit with some specific formalism has
also led to some inconsistency in the terminology.
Additional confusion is generated by studies in which
the authors quickly conclude that they are proposing
a DSR Minkowski limit whenever ``the Planck length takes the role
of an absolute scale", without verifying that the ``absolute scale"
is such to require departures from some standard special-relativistic
laws.

In light of this possibility of confusion
it is perhaps useful to contemplate explicitly
some possible roles for the Planck scale that would
indeed require a DSR Minkowski limit, and some that would not.
And let me start by observing
that if in the Minkowski limit of a given quantum-gravity theory
one had the Planck length setting
an observer-independent
minimum allowed value of wavelength, then of course
one would be dealing with a DSR Minkowski limit.
Under special-relativistic boosts wavelengths contract, and therefore
in order to enforce an observer-independent minimum-wavelength law
one should necessarily introduce departures from special relativity,
and the observer independence of the postulated new law should
allow to accommodate the departures from special relativity in
such a way that the Galilei Relativity Principle would still hold.
A role for the Planck length as observer-independent minimum wavelength
would require a modification of special relativity just like
one needs to modify Galilei Relativity in order to accommodate
a maximum-speed law (speeds transform linearly under Galilei boosts).

Similarly a DSR Minkowski limit would necessarily arise
in a quantum-gravity theory which in the Minkowski limit
predicts the existence of some absolutely fundamental particles whose
energy is constrained by an observer-independent bound ($E\leq 1/L_p$).
But of course if the Minkowski limit introduces instead a bound
on the mass (rest energy) of the particles then instead there
is no {\it a priori} reason for expecting DSR structures.
Mass is an invariant of Poincar\'e transformations, so an observer-independent
bound on mass does not necessarily affect Poincar\'e symmetry.
A useful example of the situation in which an absolute scale
does not affect symmetries is provided by the Planck constant $\hbar$
in the quantum mechanics of angular momentum. Angular momentum
transforms under space rotations, but $\hbar$ is most fundamentally
a scale affecting the square modulus of
angular momentum, $L_x^2 +L_y^2 +L_z^2$, which is an invariant under
space rotations, and in fact the scale $\hbar$ can be introduced
without affecting space-rotation symmetry~\cite{3perspe}.

For what concerns the fate of Poincar\'e symmetry in the Minkowski
limit of quantum gravity it is therefore crucial to establish
whether the Planck scale is introduced ``{\it a la} $c$"
(the speed bound introduced through $c$ required a deformation
of Galilei boosts) or is introduced ``{\it a la} $\hbar$"
(the properties of $L_x^2 +L_y^2 +L_z^2$ introduced
through $\hbar$ do not affect in any way space-rotation symmetry).
An early attempt to introduce a length scale (possibly the Planck length)
in spacetime structure in such a way that it would not require
any modification of Poincar\'e symmetry is the one of Snyder~\cite{snyder},
who indeed postulated some spacetime noncommutativity and then went
to great length to show that the system is still Poincar\'e invariant.
Some confusion may arise from the fact that in some recent papers (see, {\it e.g.},
Ref.~\cite{girelivi}) there has been some
discussion of a ``Snyder-type modification of special
relativity"\footnote{Besides the possibly confusing terminology for ``Snyder
deformations" another potential element of confusion originating
from the terminology adopted in Ref.~\cite{girelivi} (and references therein)
concerns the fact that the authors rename doubly special relativity
as ``deformed special relativity", but the name ``deformed special relativity"
had already been used in the literature~\cite{cardone}
to describe research with completely different physics motivation
and formal approach.},
as this terminology
misses the point that Snyder was trying to prove\footnote{While
Snyder should be credited for the idea of introducing the Planck scale
in spacetime structure in such a way not to affect Pincar\'e symmetry,
it is actually still unclear whether Snyder succeeded. Some of the tools
more recently developed to analyze noncommutative geometries
were not available to Snyder. Even now we only have a reliable description
of rotation and boost transformations in the Snyder spacetime,
which are indeed undeformed, whereas it is still unclear
how to properly describe translations, which are
often affected by severe ambiguities in noncommutative geometry
(see, {\it e.g.}, Ref.~\cite{alessfranc}).}
just the opposite:
an absolute length scale can be introduced without modifying
Poincar\'e symmetry.

Going back to a list of concepts which require or do not require
the concept of a DSR Minkowski limit, let me now focus on the
minimum-uncertainty intuition that is common in the quantum-gravity
literature, and is relevant for my observation concerning
the Planck-scale particle-localization limit.
If the Minkowski
limit of quantum gravity predicts an observer-independent
bound on the measurability of lengths ($\delta L \geq L_p$)
then necessarily this would have to be a DSR Minkowski limit.
Instead there is no need to introduce departures from Poincar\'e
symmetry if there is a
bound on the measurability of proper lengths (the length of
an object in its rest frame is of course a Poincar\'e-invariant
quantity).

And in closing, since there is literature on the possibility of
a maximum acceleration~\cite{caia,schu}, let me also stress that
acceleration is a Poincar\'e invariant, and therefore
a quantum-gravity theory predicting an observer-independent upper bound
on acceleration will not necessarily lead to a DSR Minkowski limit.

While this is only a very limited list of examples it should suffice
as a warning that it is not sufficient to argue that ``the Planck length
takes the role of an absolute scale" in order to provide support
for a DSR Minkowski limit. One must go through the (sometimes tedious,
but extremely valuable) exercise of deriving the explicit
form of the laws of transformation between observers, verifying
that indeed the transformation laws are Planck-scale modified.

\subsubsection*{Outlook and some challenges to the community}
The argument presented by Smolin in Ref.~\cite{leeDSRalways}
and the argument presented here combine to provide the first pieces
of a ``physics case" favouring the hypothesis of a DSR Minkowski limit,
as proposed in Ref.~\cite{gacdsr}.
It is of course important for the overall research programme to
look for other possible hints in the DSR direction, but also to look for
possible loopholes in these two arguments.

Whether or not these two arguments fully establish that a DSR Minkowski limit is
inevitable for quantum gravity, they clearly suggest that the search of formalisms
suitable for the description of a DSR Minkowski limit has been too narrow.
Some authors have identified the hypothesis of a DSR Minkowski limit
with one or another specific formalism. In turn this has led other authors
to conclude that their own preferred framework is not suitable for DSR research,
without an explicit verification of the role of the Planck scale in
relativistic transformations, but simply observing some differences from
the most popular formalisms so far adopted in the DSR literature.
An example of this type of situations is provided by research
on ``stable symmetry algebras" (see, {\it e.g.}, Ref.~\cite{vilela}),
which, understandably, has been viewed (see, {\it e.g.}, Ref.~\cite{ahluvilela})
as an alternative to the type of mathematics most popular in the DSR literature.
Still, a fully physical characterization of the laws of transformation
of {\underline{observables}} between different observers
in frameworks based on stable symmetry algebras is still missing, and
it may well be that actually some stable symmetry algebras turn out to
be useful in the description of a DSR Minkowski limit.

In developing the thesis presented in this essay I also implicitly raised
a challenge for those researchers in the quantum-gravity community who
are seeking a ``fundamental role for the Planck scale" without paying attention
to the differences between the various types of fundamental scales that
are possible in physics. For example, several papers adopt the hypothesis
that the Planck length should set the minimum allowed value for wavelengths,
but before the proposal in Ref.~\cite{gacdsr} of the idea of a DSR Minkowski limit
this minimum-wavelength studies never explored the implications for special relativity.
Similarly, there is a large literature on a vague hypothesis that the Planck
length should set the absolute limit on the measurability of lengths,
but the relevant studies often do not even provide an explicit statement concerning
whether this absolute limit applies to the measurement of proper lengths
or to the measurement of the length in any frame.
And when a limit on the measurability of lengths in any frame is assumed,
the authors often still (even now that there is some literature
on DSR Minkowski limits) do not comment on the implications for special relativity.
This point is particularly embarrassing in the case of some of the studies
based on a ``Generalized Uncertainty Principle"
attributed to string theory: nobody appears to notice that any attempt of
enforcing in full generality an uncertainty principle
of the type $\delta x \ge \frac{1}{\delta p} + \alpha \delta p$
will of course require departures from special relativity.
The opposite attitude is equally dangerous for what concerns the amount
of confusion produced in the literature: some authors, once they
have established that in a chosen framework the Planck length has the role
of ``fundamental scale", quickly jump to the conclusion that they are dealing
with a DSR Minkowski limit, whereas in order to draw such a conclusion
one should first make sure that the Planck scale affects nontrivially the laws
of transformation between inertial observers.

\bigskip
\bigskip
\bigskip

\baselineskip 12pt plus .5pt minus .5pt


\end{document}